# Boltzmann/Gibbs distribution for a two level system interacting with a thermal bath does not follow from Schrodinger equation

Iuval Clejan[†]

(919) 361-1418
clejan@mindspring.com

[†] Genetics Department, University of North Carolina at Chapel Hill, Chapel Hill NC 27599

## Abstract

In this work, we consider a 2-state quantum system interacting with a thermal reservoir. By computing the long time limit of the probability for the system to be in the ground state according to the Schrodinger/Von Neumann equation, we reach a contradiction with the prediction of equilibrium statistical mechanics. The most likely explanation is that the Schrodinger equation is incomplete as a description of such systems, because the other assumptions made herein have a wider range of experimental support.

Keywords: Boltzmann distribution, two-state system, Schrodinger equation.

## 1. Introduction

Many attempts have been made to put equilibrium statistical mechanics on a firm foundation, e.g. by derivation from the Schrodinger equation. As has been discussed previously [1], most of these attempts have either made uncontrolled perturbation expansions [2,3], invoked empirical thermodynamic arguments at some point in the "derivation" [3], or have focused on inserting stochasticity explicitly [4,5,6], which does not constitute a rigorous derivation of statistical mechanics from quantum mechanics.

Others have succeeded in deriving irreversibility, but not the Boltzmann/Gibbs equilibrium probability. One group invoked quantum chaos as being a factor responsible for irreversibility [7] in systems with even a small number of degrees of freedom, but this does not seem to be related to statistical mechanics. An example of a many body quantum system which has been demonstrated to behave irreversibly in the thermodynamic limit is the Friedrich model. In the Friedrich model, there is no distinction between system and environment. The probability of a system being in a particular state goes to zero as the number of available states around that particular state goes to infinity [3]. This result, is somewhat trivial, and does not constitute a derivation the Boltzmann/Gibbs distribution.

The simplest realistic model that may allow such a derivation would be a two state system, interacting with a thermal bath. Such a model differs from the Friedrich model in that the Hilbert space considered is a tensor product space of the system and the environment, and the probability of the system being in any of its two states can be computed and compared to the Boltzmann/Gibbs form. We are not aware of any peer reviewed work which treats this model without artificially introducing stochasticity. One non-peer reviewed treatment which does not introduce stochasticity [8] appears to contain mathematical errors and questionable approximations, and we were unable to replicate its results.

The efforts to derive the Boltzmann/Gibbs equilibrium probability distribution starting with the Schrodinger (or equivalently for the density matrix, starting with the Von Neumann) equation have not been fruitful, not because they haven't found the right method. As we will show, for most n state systems interacting through any interaction with a thermal bath and evolving through the Schrodinger equation, the Boltzmann/Gibbs form is not achieved.

## 2. Long Time Averaged System Density Matrix

Consider a 2 level system interacting with a thermal reservoir. We assume the thermal reservoir is initially in equilibrium at inverse temperature β, and described by a density matrix of the form:

$$\rho_R = \text{Exp}(-\beta H_R)/Z \tag{1}$$

where Z is the partition function of the reservoir. We will attempt a sort of mathematical induction, or rather the inductive step of a mathematical induction—that is we will assume that the reservoir has equilibrated before it interacts with our two level system, and see if the two level system can equilibrate to the temperature of the reservoir (in the limit of small interaction strength), as statistical mechanics predicts. We will see that the two level system will not equilibrate to the temperature of the reservoir.

Assume, for simplicity that the system is prepared in its ground state, so that the initial density matrix of the composite (system plus reservoir) is given by:

$$\rho(0) = \sum_{j=1}^{N} A_j |0j\rangle\langle 0j|, \tag{1}$$

where

$$A_j = \text{Exp}(-\beta E_j)/Z \tag{3}$$

and |ij> denotes the state where the system is in its state |i> (i=0 or 1) and the environment is in its state |j> (j ranges from 1 to N, with N large).

Assume that the system interacts with the reservoir through a microscopic, temperature independent interaction, and that the non-interacting eigenfunctions of the composite are related to the interacting eigenfunctions through the matrix T:

$$|l\rangle = \sum_{ij} T_{l(ij)} |ij\rangle \tag{4}$$

Now allow the composite to evolve according to the Schrodinger (or equivalently, Von Neumann) equation. This evolution is most easily written in the basis of interacting eigenfunctions of the composite, which have a simple evolution:

$$\rho(t) = \sum_{jlm} A_j T^*_{l(0j)} T_{m(0j)} |l(t)\rangle\langle m(t)|$$
$$= \sum_{jlm} \mathrm{Exp}[i(\omega_l - \omega_m)t] A_j T^*_{l(0j)} T_{m(0j)} |l\rangle\langle m| \qquad (5)$$

In order to obtain the probability for the system to be in the ground state, we need not the composite density matrix, but the system density matrix (SDM). The system density matrix is obtained by tracing over all the reservoir states:

$$\rho_{S_{no}} = \sum_{jlmk} \mathrm{Exp}[i(\omega_l - \omega_m)t] A_j T^*_{l(0j)} T_{m(0j)} T_{l(nk)} T^*_{m(ok)} \qquad (6)$$

We now specialize to the case n=o=0 (which gives the probability of the system being in the ground state), and note that if equilibrium is to be achieved for large number of bath particles N and large volume V, the only terms which do not cancel each other are those for which $\omega_l=\omega_m$, giving

$$P_0 \equiv \rho_{S_{00}} = \sum A_j f_j \qquad (7)$$

where

$$f_j = \sum_{lmk} T^*_{l(0j)} T_{m(0j)} T_{l(0k)} T^*_{m(0k)} \qquad (8a)$$

with $\omega_l=\omega_m$. Another way to achieve $\omega_l=\omega_m$ is to take a time average and note that the only terms that survive the time average in eqn (6) are the ones for which $\omega_l=\omega_m$, since other terms are oscillatory and their contribution to the integral goes to zero for large averaging time when they are divided by the averaging time. Still keeping N and V large but finite, we also know that the density of states increases exponentially with N and so the sum can be approximated by an integral:

$$f(x) = \iiint dy\,dw\,dz\, T^*_{y(0x)} T_{z(0x)} T_{y(0w)} T^*_{z(0w)} \Omega(x) \qquad (8b)$$

where $\Omega(x)$ is the (many particle) density of states of the composite. Note that all the temperature dependence of $P_0$ is contained in A(x) (the continuum version of $A_j$), and f(x) is temperature independent. Actually, the precise form of f and the validity of the constraint $\omega_l=\omega_m$ is not important as long as f is temperature and time independent.
We can rewrite eqn (7) as:

$$P_0 = \frac{1}{Z(\beta)} \int_0^\infty \mathrm{Exp}[-\beta x] f(x) dx \qquad (9)$$

where eqn (3) for *A(x)* was substituted into eqn (9). On the other hand, equilibrium statistical mechanics predicts that the probability of the system being in the ground state, once the interaction strength approaches zero, for large but finite N and V is:

$$P_0 = \frac{1}{1 + Exp[-\beta\delta]} \tag{11}$$

where δ is the energy spacing between the two states of the system. Equating (9) and (11), we obtain:

$$\int_0^\infty Exp[-\beta x] \bar{f}(x) dx = \frac{\bar{Z}(\beta)}{(1 + Exp[-\beta\delta])} \tag{12}$$

where the bar above f and Z denotes a renormalization. Note that both f(x) and Z(β) increase with N and V in the same manner (in order for $P_0$ to be independent of N and V), and so if we are concerned about divergences in equation 12, we can normalize both f(x) and Z(β) by a common (temperature independent) factor to give $\bar{f}(x)$ and $\bar{Z}(\beta)$. It can be easily seen that $\bar{f}(x)$, since it's independent of temperature and by the uniqueness of the Laplace transform, is the inverse Laplace transform of the function on the RHS of (12). However, there is no function whose inverse Laplace transform is the RHS of (12). This is because the inverse Laplace transform, if it existed would be the sum of the residues at the poles of the RHS of (12). The poles occur on the imaginary β axis at β'=i(2n+1)π/δ (n any integer) with a residue of $Exp(\beta')\bar{Z}(\beta')/\delta$. The only way the sum over residues can converge is if Z(β) has a zero at each of these poles, or if Z(β') decreases with increasing n such that the sum is convergent, or if there is cancellation of terms due to some symmetry. There is no reason why any of these possibilities should be true in general and indeed they are not true (as can be easily checked) for the partition function of a gas of non-interacting particles, whether they are fermions, bosons or distinguishable classical particles. Ising lattice systems have a discrete spectrum and do no concern us here. Note also that we didn't explicitly use the fact that the interaction strength goes to zero, but we used it implicitly by assuming that the equilibrium temperature of the composite in that case is the same as the initial temperature before the interaction was turned on.

## 3. Discussion

Thus we have arrived at a contradiction by assuming that:

1. The environment is initially in equilibrium at inverse temperature β.
2. The environment has a density matrix described by the Boltzmann form.
3. The interaction between system and environment is temperature independent.
4. The Schrodinger equation describes the evolution of the composite.
5. The final temperature of the environment is the same as the initial temperature because the interaction strength goes to zero.

6. The equilibrium probability of the system to be in its ground state is given by the Boltzmann/Gibbs form, with the same temperature as the environment.

Assumption 1 is an ansatz. Assumption 2, although never rigorously tested experimentally (because the environment has too many states and not each one can be measured), has plenty of experimental and theoretical support, and is one of the foundations of equilibrium statistical mechanics. Assumption 3 is a reflection of the microscopic nature of the interaction. Mesoscopic or macroscopic interactions can be heuristically treated as temperature dependent, but microscopic interactions (e.g. electromagnetic) have nothing to do with temperature. Assumption 5 is a way of saying that the environment is large enough to not be perturbed by interaction with the system. As for assumption 6, experimental validations of the Boltzmann/Gibbs distribution for a two state system at thermal equilibrium are abundant (see for example [9] for a paramagnetic salt of spin ½). In addition, the Boltzmann/Gibbs form of the density matrix is known to be entropy maximizing for a fixed temperature, so if it is not achieved, the composite violates the second law of thermodynamics, which has also plenty of experimental backing. We are left with assumption 4 and forced to conclude that the Schrodinger equation is either incomplete or wrong for a two state system interacting with a thermal bath.

It is easily seen that the above argument can be generalized to an n level system, with eqn (12) having a more complicated sum in the denominator on the RHS.

## 4. Conclusions

We have shown an inconsistency between the Schrodinger equation and the (experimentally verified) predictions of equilibrium statistical mechanics for a two level system interacting with a thermal reservoir. It seems remarkable that an inconsistency between the Schrodinger equation and experiment should not have been observed for so many years, until now. However, the experimental validation of the Schrodinger equation has been either for statics (calculation of energies), or for dynamics of simple (few body) systems. There have not been experimental validations of the non perturbative Schrodinger equation for dynamics of many body systems, which is what we are considering in this work. At the same time, there have been no numerical solutions of the Schrodinger equation for many body systems such as the one considered in this work, which had they been computed, would have found the inconsistency with statistical mechanics. It is beyond our scope to speculate on why upon going from few body systems to many body systems, the Schrodinger equation should cease to be valid.

## Acknowledgements

I wish to thank Dr. Evelyn Wright and Dr. Linda Reichl for useful discussions and suggestions, and Dr. Ganpathy Murthy for useful suggestions.